# Structural and Dielectric Properties of (La, Nd) (Mg$_{1/2}$Ti$_{1/2}$)O$_3$ Perovskites

Kouros Khamoushi and Eero Arola

*Department of Electrical Energy Engineering, Tampere University of Technology,
P.O. Box 692, 33101 Tampere, Finland. E-mail: kouros.khamoushi@tut.fi*

**Abstract**
Using the high-resolution x-ray diffraction (XRD) analysis, scanning electron microscopy (SEM), and the temperature-dependent microwave resonator characterization, structural properties, phase assemblage and dielectric properties of La(Mg$_{1/2}$Ti$_{1/2}$)O$_3$ (LMT) and Nd(Mg$_{1/2}$Ti$_{1/2}$)O$_3$ (NMT) ceramics prepared via the mixed oxide route were investigated in this study. Single-phase ceramics were synthesized for both LMT and NMT at sintering temperatures from 1250ºC to 1675ºC. On the basis of the XRD analysis we have found that the LMT and NMT compounds have cubic and monoclinic crystal structures, respectively. We have also observed that the relative densities of LMT and NMT vary between about 93 and 99% of the theoretical density, depending on the sintering temperature. Finally, concerning the dielectric properties of the microwave resonators made of LMT and NMT compounds we have measured their temperature coefficient of the resonant frequency ($\tau_f$) and the quality factor ($Q$). It is interesting to notice that $\tau_f$ in the case of the NMT compound (-16 ppm K$^{-1}$) is essentially smaller than in the case of the LMT compound (-72 ppm K$^{-1}$), therefore proving a better stability against temperature variations in the NMT based resonators. On the other hand, the $Q$ values are very similar, being 34000 at the resonance frequency of 8.07 GHz and 38000 at the resonance frequency of 9.76 GHz in the LMT and NMT cases, respectively.
**Keywords:** dielectric properties; microstructure-final: grain growth; single phase Ln(Mg$_{1/2}$Ti$_{1/2}$)O$_3$ (Ln=Nd, La); Lanthanum magnesium titanium oxide.

## 1. Introduction

A small ceramic component made of a dielectric material is fundamental to the operation of filters and oscillators in several microwave systems, such as satellite TV receivers, military radar systems, and mobile communications.

In microwave communications, dielectric resonator filters are used to discriminate between expected and unwanted signal frequencies in the transmitted and received signal. When the expected frequency is extracted and detected it is also necessary to maintain a strong signal level. It is also critical that the expected signal frequencies are not affected by seasonal temperature changes. The resonator materials for practical applications should have certain key properties.

A high relative dielectric constant is needed so that the materials can be miniaturized and a high quality factor (Q) is needed for an improved selectivity. The quality factor of the microwave resonator generally describes its capability of storing electromagnetic energy at the resonant frequency $f_0$ with as small as possible energy dissipation (losses) in the dielectric material. In the simplest experimental setup the various power losses (conduction, dielectric, and radiation losses) of the resonant cavity and the external load are not been separated from each other. In this case, one defines the so called *loaded quality factor* $Q_L$ which can directly be measured from the resonance peak of the transmitted signal from the combined system of the resonant cavity and the external load as

$$Q_L \equiv \frac{f_0}{\Delta f}, \quad (1)$$

where $f_0$ is the resonance frequency of the dielectric resonator and $\Delta f$ is the full width at half maximum of the resonance peak (FWHM), i.e. the band width at 3 dB damping of the power signal.

In addition to a large quality factor, a low temperature variation of the material's resonant frequency is also required in order that the microwave circuits remain stable. Everything from the electromagnetic properties to the microstructure of the material is important for the final result.

In order to decrease the size, weight and cost of microelectronic devices principles of tuning the temperature coefficient of resonant frequency ($\tau_f$) in complex perovskites have already been established. Colla *et al*. [1,2] and Reaney *et al*. [3] have shown that the temperature coefficient of dielectric constant ($\tau_\varepsilon$) in Ba- and Sr-based complex perovskites is fundamentally related to the onset and degree of octahedral tilting. Moreover, it can be tuned through ±300 ppm K$^{-1}$ without significantly altering Q or $\varepsilon_r$ by manipulating the perovskite tolerance factor, *t*,

$$t = (R_A + R_O)/\sqrt{2}(R_B + R_O) \quad (2)$$

between 0.93-1.01 where $R_A$, $R_B$, and $R_O$ are the radii of the ions in the perovskite (ABO$_3$) structure. Reducing *t* results in the onset of octahedral tilt transitions. The relationship between $\tau_\varepsilon$ and $\tau_f$ is:

$$\tau_f = -\left(\tau_\varepsilon/2 + \alpha_L\right), \quad (3)$$

where $\alpha_L$ is the coefficient of linear thermal expansion ($\approx$10 ppm K$^{-1}$ for perovskites).

La(Mg$_{1/2}$Ti$_{1/2}$)O$_3$ (LMT) is a perovskite which forms in the cubic system, although its exact crystal structure remains unclear. Its tolerance factor $t = 0.95$, which, according to the work of Lee et al. [4], indicates the presence of both in-phase and anti-phase tilting of oxygen octahedra. The effects of this tilting have been observed by XRD in the form of ½(111) ordering at 1600ºC. Superlattice reflections corresponding to anti-phase and in-phase tilting have been observed. Systematic cation displacement was also detected by the appearance of ½(210) reflections, and ½(111) reflections were observed and ascribed to the doubling of the unit cell caused by the B-site cation ordering. Based on the tilting of the oxygen octahedra two possible space groups are characterizing the LMT crystal, namely Pbmn or P2$_1$/n. However Walsenburg photographs or TEM work would be required to unambiguously establish the structure.

The Nd-analogue, Nd(Mg$_{1/2}$Ti$_{1/2}$O$_3$) (NMT), has a tolerance factor of $t=0.9154$, which would suggest a highly tilted oxygen octahedral structure on the verge of perovskite/ilmenite stability, where the tilting occurs both in-phase and anti-phase. Structurally, tilting of octahedra has similar effects as cation ordering in 1:1 type complex perovskites. Both result in doubled unit cells [5] however, cation ordering also reduces the space available where the A-site species can move around. The structure and microwave dielectric properties of Nd(Zn$_{1/2}$Ti$_{1/2}$)O$_3$ (NMT) and La(Mg$_{1/2}$Ti$_{1/2}$)O$_3$ (LMT) were investigated in this research.

## 2. Experimental Procedure

In the following only a brief account of the material processing details for the LMT and NMT compounds is given. Further details can be found in Ref. [6].
In this work, conventional mixed oxide powder processing techniques were used. Starting materials included La$_2$O$_3$ (99.9% Meldform Rare Earths, U.K.), Nd$_2$O$_3$ (99.9% Meldform Rare Earths, U.K.), TiO$_2$ (99.8% Alfa Aesar, U.K.), and ZnO (99.9% Elementies Specialties, U.K.), (MgCO$_3$)$_4$Mg(OH)$_2$5H$_2$O (99% Aldrich Chemical company, Inc, USA). For both LMT and NMT, the rare-earth oxide was first purposely hydrated in distilled water to form either La(OH)$_3$ or Nd(OH)$_3$. These hydrates were then used in the subsequent processing procedure, which involved milling stoichiometric amounts of powders together in a porcelain mill pot partly filled with ZrO$_2$ media and distilled water for four hours.

A small amount (1 wt%) of Dispex A40 (Allied Colloids, Bradford, U.K.) was added as a deflocculant. The slurries were then dried overnight at 80°C. Dried powders were subsequently granulated with a mortar and pestle and sieved to less than 250 μm in size. Calcination was then achieved in a two-stage process described in the abovementioned reference [6].
These slurries were then dried and granulated as above and subsequently pressed (125 MPa) into cylindrical pellets 10 mm in diameter and 3 mm thick. Sintering was conducted in closed alumina boats for 6 hours at temperatures ranging from 1400°C to 1675°C for LMT and NMT. Pellets were weighed before and after sintering to quantify the degree of mass loss.
Phase assemblages and crystal structures were investigated by the scanning electron microscopy (JSM 6300, Jeol, Tokyo) and x-ray diffraction (D50000, Siemens, Germany, Cu Kα radiation) facilities, respectively.

## 3. Result and discussion

Figure 1 shows the density of LMT and NMT as a function of sintering temperature. We can see that the density of LMT increases with temperature and reaches to a 87% of theoretical density at 1250°C, then it increasing to 98% of theoretical density at 1500°C. The density of NMT increases with temperature and reaches to a 88% of theoretical density at 1250°C, then it increasing to 94% of theoretical density at 1500°C.

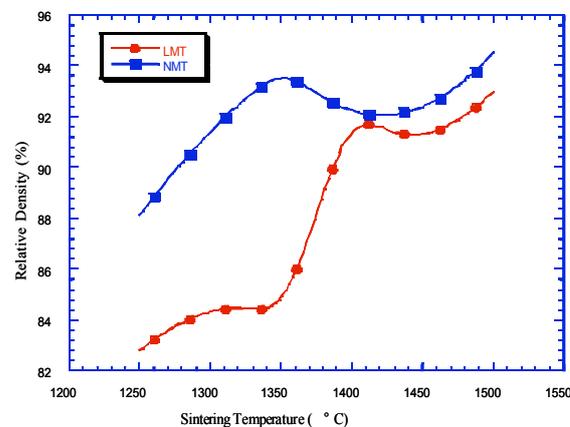

**Fig. 1** Relative density of LMT and NMT pellets after sintering for 6 hours at 1250-1500 ºC.

Figure 2 shows the XRD patterns of the LMT samples sintered at 1400-1675ºC temperature. Clearly all the samples show perovskite crystal structure in the XRD diagrams. The Crystallographica Search-Match program indicates that crystal structures of these samples are cubic with a space group of Pa3.

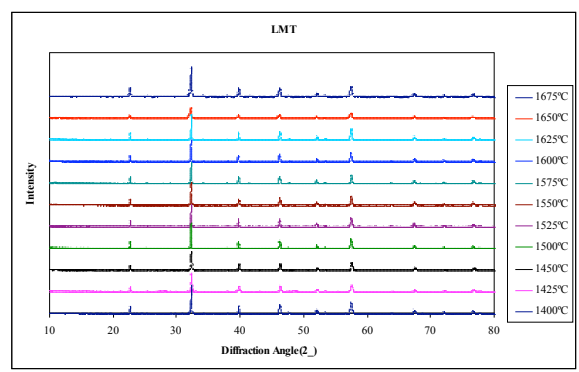

**Fig. 2** X-ray diffraction traces of LMT at 1400-1675ºC.

Figure 3 illustrates XRD patterns of the NMT samples sintered at 1400-1675ºC temperatures. All the samples show again perovskite phase in the XRD diagrams. Crystallographica Search-Match program indicates that the crystal structures of these samples are monoclinic.

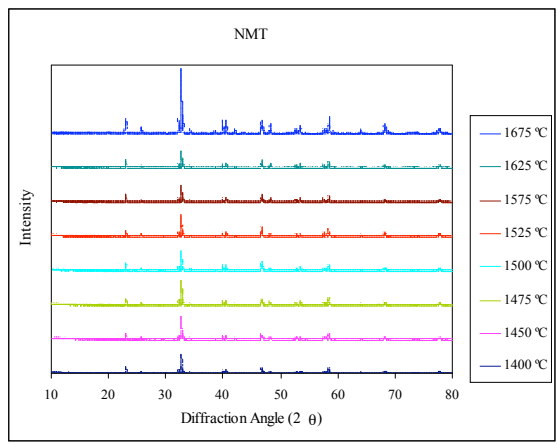

**Fig. 3** X-ray diffraction traces of NMT at 1400-1675ºC.

Figures 4 and 5 show the SEM secondary electron images of $La(Mg_{1/2}Ti_{1/2})O_3$ (LMT) and $Nd(Mg_{1/2}Ti_{1/2})O_3$ (NMT) compounds sintered at 1600°C, respectively. Each pellet is a single phase dense ceramic with a grain size of approximately 1-5 $\mu$m. From experimental viewpoint material with a different phase must show a different contrast in the SEM image. However, backscattered electron images of these samples did not show any differences in the contrast.

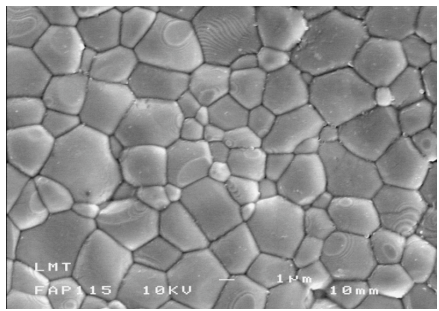

**Fig. 4** A SEM image of a pellet of LMT sintered at 1600ºC.

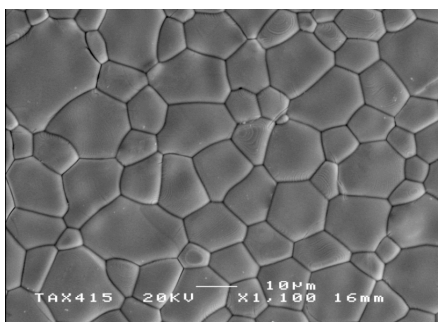

**Fig. 5** A SEM image of a pellet of NMT sintered at 1600ºC.

### 3.1. Structure of LMT

The X-ray pattern for the LMT powder is shown in Fig. 6. The diffraction pattern indexing has been carried out by using the Magnesium Lanthanum Titanium Oxide Powder Diffraction File (PDF) card number 49-242 with a single phase cubic crystal structure with the lattice parameter $a \approx 0.39195$ nm in connection with the Crystallogaphic Search-Match program. Furthermore, the XRD peaks of the powder sample did not show any broadening or splitting, therefore indicating no disorder or symmetry distortion (reduction). All the XRD diagrams taken from the calcined as well as the sintered pellet samples were similar.

However, in order to avoid the ambiguity in determining the exact crystal structure and space group of these LMT samples, the high-resolution transmission electron microscopy (HRTEM) and Rietveld refinement by using General Structure Analysis System (GSAS) should be performed.

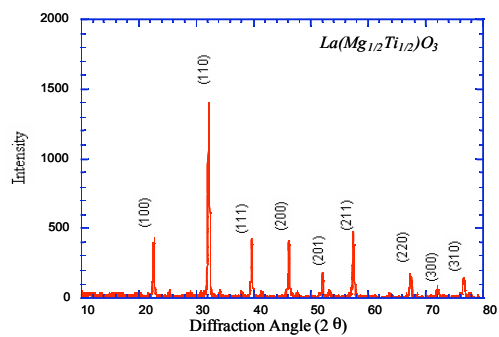

**Fig. 6** X-ray diffraction pattern of the LMT powder after calcination at 1600°C. All the peaks have been indexed according to the PDF card number 49-242 and a cubic perovskite unit cell with the unit cell parameter $a \approx 0.39195$ nm.

### 3.2. Structure of NMT

The X-ray pattern for the NMT powder is shown in Fig. 7. It is indexed according to the Neodymium Magnesium Titanium Oxide PDF card number 77-2426 with a single phase monoclinic crystal structure with the following lattice parameters: $a \approx 0.27330$ nm, $b \approx 0.27953$ nm, and $c \approx 0.38840$ nm within the $P2_{1/n}$ space group. As in the LMT case, the XRD peaks of the powder samples did not show any broadening or splitting. Furthermore, all the XRD patterns of the calcined and sintered pellets were similar. On the other hand as in the LMT case, to avoid ambiguity in determination of the exact crystal structure and space group of the NMT samples, the HRTEM and Rietveld refinement techniques are required.

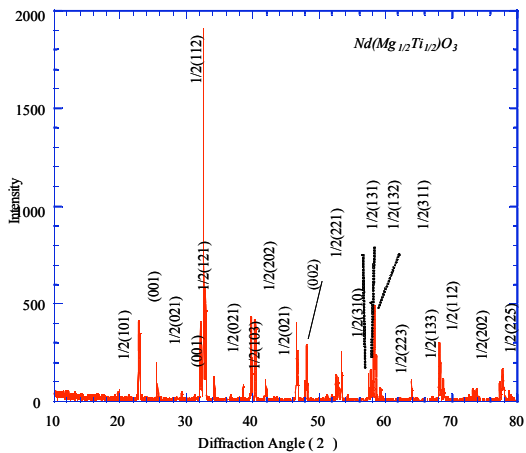

**Fig. 7** X-ray diffraction pattern of the NMT powder after calcination at 1675°C. All the peaks have been indexed according to the PDF card number 77-2426 and monoclinic perovskite unit cell.

### 3.3. Dielectric properties of LMT and NMT

We finally briefly discuss the dielectric properties of the LMT and NMT compounds in terms of the temperature coefficient of the microwave resonant frequency ($\tau_f$) and the quality factor ($Q$) at the resonant frequency.

Interestingly, our experiments show that the temperature coefficient of the resonant frequency $\tau_f$ for the LMT based resonator is -72 ppm $K^{-1}$, while that in the case of the NMT based resonator becomes essentially smaller, namely -16 ppm $K^{-1}$. Therefore, in comparison to the LMT compound, the NMT compound used in the microwave resonators will clearly provide a better stability against temperature variations.

However, considering our experiments on the quality factor Q of the microwave resonators, the LMT and NMT compounds, used as resonator materials, result in very similar values for Q, namely 34000 at the resonance frequency of 8.07 GHz and 38000 at the resonance frequency of 9.76 GHz, respectively.

### 4. Conclusions

Structural and dielectric properties of La(Mg$_{1/2}$Ti$_{1/2}$)O$_3$ (LMT) and Nd(Mg$_{1/2}$Ti$_{1/2}$)O$_3$ (NMT) compounds were investigated. Through X-ray diffraction experiments we found out that the LMT and NMT compounds possess cubic and monoclinic crystal structures, respectively.

We have also observed that the relative densities of LMT and NMT vary between about 93 and 99% of the theoretical density, depending on the sintering temperature. Finally, concerning the dielectric properties of the microwave resonators made of LMT and NMT compounds we have measured their temperature coefficient of the resonant frequency ($\tau_f$) and the quality factor ($Q$). Interestingly, we notice that $\tau_f$ in the case of the NMT compound is essentially smaller (-16 ppm $K^{-1}$) than in the case of the LMT compound (-72 ppm $K^{-1}$), therefore proving a better stability against temperature variations in the NMT based resonators. On the other hand, the $Q$ values are very similar, being 34000 and 38000, in the LMT and NMT cases, respectively.

On the basis of our study we conclude that both LMT and NMT compositions have promising dielectric properties showing therefore a great potential as dielectric filters in the mobile microwave telecommunication technologies.

### Acknowledgments

One of us (K. K.) is thankful to Profs. Neil Alford and Rick Ubic for their help and guidance at London South Bank University and Queen Mary University of London, respectively, where this research has been conducted.